\documentclass[aps, amsmath, twocolumn, amssymb,groupedaddress]{revtex4}
\usepackage{graphicx}
\usepackage{gensymb}
\pdfoutput=1
\usepackage{amsmath}
\usepackage{color}
\usepackage{float}
\usepackage[english]{babel}
\usepackage{lipsum}
\usepackage{color,soul}

\begin{document}
\title{Comment on ``Perfect Andreev reflection due to the Klein paradox in a topological superconducting state (Nature \textbf{570}, 344 (2019))"}

\author{Shekhar Das$^1$}
\author{Kurt Gloos$^2$}
\author{Yu. G. Naidyuk$^3$}
\email{naidyuk@ilt.kharkov.ua}
\author{Goutam Sheet$^1$}
\email{goutam@iisermohali.ac.in}
\affiliation{$^1$Department of Physical Sciences,  
Indian Institute of Science Education and Research Mohali,
Sector 81, S. A. S. Nagar, Manauli, PO: 140306, India}
\affiliation{$^2$Wihuri Physical Laboratory, Department of Physics and Astronomy, University of Turku, FIN-20014 Turku, Finland}
\affiliation{$^3$B. Verkin Institute for Low Temperature Physics Engineering NAS of Ukraine, Kharkiv, Ukraine}

\maketitle

In a recent publication \cite{1}, Lee $et$ $al.$ discussed experimental observation of Klein tunneling into a proximity-induced topological superconducting state of SmB$_{6}$ through the measurement of Andreev reflection with low-bias conductance doubling across point-contact junctions between sharp Pt-Ir tips and  SmB$_{6}$/YB$_{6}$ heterostructures. However, the interpretation of the presented point-contact data is rather ambiguous because observing a low-bias conductance enhancement by a factor of 2 is not special for Klein tunneling into topological superconductors. Such an enhanced conductance can be observed in point contacts between all types of superconductors and normal metals when the contacts are neither in the ballistic nor in the diffusive regimes of mesoscopic transport, but in a dissipative thermal regime. The thermal regime is expected for the point contacts presented in \cite{1} primarily because of the anticipated very short electron mean free path of SmB$_6$ -- we will discuss this in more detail later.

First we note that one common (but not necessary) feature that can be used to identify $dI/dV$ spectra of point contacts in thermal regime is the appearance of conductance dips near the superconducting gap \cite{2,3}. Such dips may be sharp or shallow, and cannot be described by the Blonder-Tinkham-Klapwijk (BTK) theory, a modified version of which has been used by the authors of \cite{1}.  Most of the spectra presented in \cite{1} indeed show such conductance dips. In those cases, the variation of the conductance at low-bias is not entirely due to Andreev reflection. It also contains the thermal or so-called Maxwell component of the point contact resistance. This component causes a conductance dip when the electrical current through the contact exceeds a critical value \cite{2,3}. To note, an alternative explanation for the enhanced conductance and the dips, which can probably be excluded here for obvious reasons, could be proximity-induced superconductivity in the normal electrode as discussed in Ref. 29 of \cite{1}.

\begin{figure}[h!]
	\centering
	\includegraphics[width=.5\textwidth]{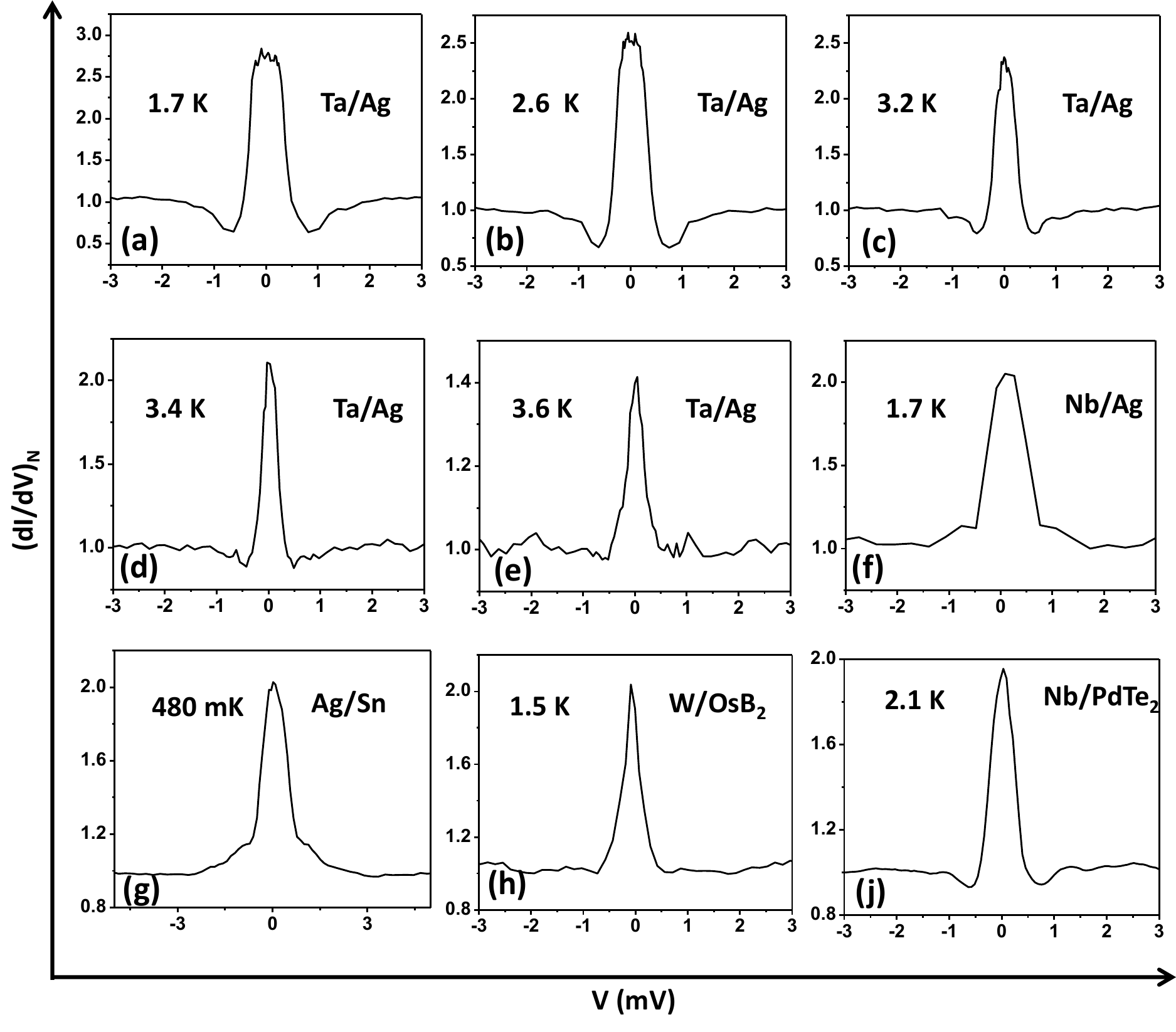}
\caption{(a-e) Thermal limit point contact spectra between superconducting Ta and Ag at different temperatures. Thermal limit point contact spectra between superconducting (f) Nb and Ag, (g) Sn and Ag, (h) OsB$_{2}$ and W, (j) PdTe$_{2}$ and Nb.}
\label{Figure1}
\end{figure}

The authors of \cite{1} claim that there are only two  reports in the existing literature showing conductance enhancement greater than 1.9.  This is incorrect, as many more such spectra have been reported in a number of places\cite{2, 3, Walti, Gloos, Daghero}. In fact, such spectra are encountered frequently in point contact spectroscopy experiments involving superconductors. For example, in Figure \ref{Figure1} (a-e), we show some representative spectra with similar features in thermal point contacts on conventional BCS superconductors. Most of the times such spectra are discarded (and not published) because they do not provide any spectroscopic information, the corresponding contacts being far away from the spectroscopic (ballistic) regime. To note, in non-ballistic point contacts with heavy-fermion superconductors \cite{Walti,Gloos}, the contacts can display a low-bias conductance enhanced by a factor of 2, or even more than 2, indicating dominant contribution of Maxwell’s resistance and a negligible contribution of Andreev reflection \cite{Gloos}. Furthermore, the low-bias enhancement in such cases may be even more pronounced at lower temperatures because the critical current decreases with increasing temperature. For example, in Figure 1 (a-e) we show how temperature causes broadening of spectra for thermal contacts between the conventional superconductor Ta and Ag. In \cite{1}, the authors did not report any spectrum below 2 K and hence it is not clear whether at lower temperatures the zero-bias conductance would exceed 2 and make their interpretation even more unrealistic.

         We also note that the conductance enhancement observed in \cite{1} may not be related to a strongly reduced small $Z$ parameter of normal reflection (as claimed by the authors). This is because even for ballistic point contacts the $Z$ parameter depends only weakly, if at all, on Fermi velocity mismatch \cite{6}-- the dominant contribution comes from the physical character of the interface.  In Figure \ref{Figure1} (h,j) we present two spectra from point contacts between materials with significantly different Fermi velocities (Nb/PdTe$_{2}$ and W/OsB$_{2}$) - but, the low-bias conductance is seen to be enhanced by a factor of 2. The same can also be seen with even more complex superconductors\cite{Walti, Gloos, Daghero}. 

The presented analysis of the data in \cite{1} should also be considered critically. First, the presented BTK fits are, in fact, far from perfect. The discrepancy is actually masked by the big size of the experimental data points. Second, the authors report that the point-contact resistance for the presented data amount to only a few ohms. The same authors had demonstrated earlier that their 25 nm thick SmB$_{6}$ films had resistance greater than 600 Ohm at 1 K \cite{7}. This indicates a short electron mean free path and a high electrical resistivity. Thus, the point contacts shown in \cite{1} must be of large size. The latter is confirmed by the authors themselves when they write in the supplementary information -- ``Extended Data Fig. 3b shows the top view of an Au-SmB$_{6}$/YB$_{6}$ structure with a circular junction (diameter, 10 $\mu$m)". In any case, for such huge size junctions realization of ballistic condition is impossible. To note, the spectral features reported with Pt-Ir tips show striking similarities with those with Au tips indicating the contacts with Pt-Ir are also large and therefore away from the ballistic regime. Third, according to Extended Data Fig. 4 in \cite{1}, the resistance of their films increases only 4 times at low temperatures, indicating that the SmB$_{6}$ films are not very good insulators and they have significantly larger defect density compared to SmB$_{6}$ single crystals, where the resistance at low temperatures can increase by more than four orders of magnitude \cite{7}. To re-emphasize, such imperfections in SmB$_{6}$ films are expected to result in short electron mean free path, which together with large point-contact size makes the ballistic condition impossible for the presented point contacts in \cite{1}.

	Finally, according to Szabo $et$ $al.$ \cite{8}, the coherence length in YB$_{6}$ is approximately 30 nm. In \cite{1}, the induced gap in SmB$_{6}$ probed by a point-contact on top of the film must decrease on a similar length scale with increasing thickness of the SmB$_{6}$ film. However, according to Figs. 1 and 2 in \cite{1}, the thinnest 10 nm SmB$_{6}$ film (Fig.2b) has a smaller gap of 0.59 meV than the films with thickness 20 and 30 nm which both have gaps of about 0.75 meV (Fig.1 c,d) – this is unphysical and provides very strong support for the main argument of this comment that the authors have erroneously extracted spectroscopic information from point contacts far away from the regimes of transport where energy resolved spectroscopy can be performed.

Therefore, considering the discussion above, it is most rational to conclude that the claim of perfect Andreev refection due to Klein tunneling by the authors of \cite{1} is ambiguous.

\end{document}